\documentclass[aps,prc,twocolumn,showpacs,superscriptaddress]{revtex4-2}
\usepackage{natbib}
\usepackage{graphicx}
\usepackage{color}
\usepackage{physics}
\usepackage{amsfonts}
\usepackage{tabularx}
\newcolumntype{Y}{>{\centering\arraybackslash}X}
\begin{document}
\bibliographystyle{revtex}
\title{Bulk and spectroscopic nuclear properties within an  {\it ab initio} renormalized random-phase approximation  framework}

\vspace{0.5cm}
\author{R. Folprecht}
\affiliation{Institute of Particle and Nuclear Physics, Faculty of Mathematics and Physics,  Charles University, V Hole\v sovi\v ck\'ach 2, 180 00 Prague, Czech Republic } 
\author{F. Knapp}   
\affiliation{Institute of Particle and Nuclear Physics, Faculty of Mathematics and Physics,  Charles University, V Hole\v sovi\v ck\'ach 2, 180 00 Prague, Czech Republic } 
\author{G. De Gregorio}
\affiliation{Dipartimento di Matematica e Fisica, Universit$\grave{a}$ degli Studi della Campania "Luigi Vanvitelli",
viale Abramo Lincoln 5, I-81100 Caserta, Italy}
\affiliation{Istituto Nazionale di Fisica Nucleare, Complesso Universitario di Monte S. Angelo, Via Cintia, I-80126 Napoli, Italy}
\author{R. Mancino}
\affiliation{Institute of Particle and Nuclear Physics, Faculty of Mathematics and Physics,  Charles University, V Hole\v sovi\v ck\'ach 2, 180 00 Prague, Czech Republic }
\author{P. Vesel\'y} 
\affiliation{Nuclear Physics Institute,
Czech Academy of Sciences, 250 68 \v Re\v z, Czech Republic} 
\author{N. Lo Iudice}
\affiliation{Dipartimento di Fisica, 
Universit$\grave{a}$ di Napoli Federico II, 80126 Napoli, Italy} 

\date{\today}

\begin{abstract}
A modern chiral potential incorporating the three-body force is adopted to investigate bulk properties, spectra, and nuclear responses of closed-(sub)shell nuclei throughout the nuclear chart within a particle-hole (p-h) renormalized random-phase approximation (RRPA) scheme using a Hartree-Fock (HF) single-particle  basis. Our analysis shows that all instabilities induced by the quasiboson approximation (QBA) underlying RPA are removed and an overall better consistency with the experiments is achieved for all observables of the investigated nuclei.
The residual discrepancies point out the need of going beyond the p-h space. 
\end{abstract}

\maketitle

{\it Introduction.} 
RPA is one of the most widely adopted methods devoted to the study of nuclear spectroscopy. 
It can be derived from the equation of motion method \cite {rowe10} or from the linear response theory \cite {ring04}.

It is an essential tool for investigating nuclear spectra and response functions, including pygmy and giant resonances, as well as weak-interaction processes that are of significant importance in astrophysical contexts.

 RPA was implemented within many different frameworks, some using non-relativistic (e.g. Refs. \cite{Bend03,Roca18}) or relativistic (e.g. Refs. \cite{Vret05,Paar08}) density functionals, other exploiting the Green's function formalism (e.g. Refs. \cite{Lenske90,Lit23}).   
 The latter approach allows one to go beyond the harmonic approximation by coupling the RPA modes to complex configurations.
 Other extensions using Skyrme forces were proposed \cite{Gamba15}.
The investigations based on modern realistic potentials are few and rather recent \cite{Paar06,Papa10,Herg11,Wu18,Rai19,Beau23}.

RPA is known to be an extension of Tamm-Dancoff approximation (TDA). In both approaches, the eigenvalue problem is formulated within a p-h or two quasiparticle (2qp) spaces. In TDA, the ground state is the  HF vacuum. In RPA, it is assumed to be correlated in the formal derivation of the equations of motion, but is replaced by an uncorrelated HF wavefunction in the actual calculation of the RPA matrix elements. This simplification, known as quasi-boson approximation (QBA), induces instabilities at low excitation energies.

Most of the recipes for obviating this shortcoming were enumerated and developed long ago  by Rowe \cite{Rowe68,Rowe68a}. A more updated list of papers devoted to this task is presented and discussed in a recent  review \cite{Schu21}. They consist in reintroducing the correlations into the ground state without spoiling the simple structure of the RPA eigenvalue equations. Few investigations were conducted along this line since then \cite{ELLIS87,KARA93,CATA94}. An important extension was achieved and applied to metal clusters in Refs. \cite{Cat96,Cat98}. 

The method so reformulated was applied to nuclei by Papakonstantiou et al. \cite{Papa07} using the UCOM potential derived from the Argonne V18 nucleon-nucleon (NN) interaction through a unitary transformation. The effect of the ground-state correlations on several multipole responses resulted to be modest.


Several alternative approaches accounting for  ground-state correlations are available. We mention an extension of TDA \cite{Mina16}, a time-dependent density matrix (TDDM) formalism \cite{Bart21}, a many-body perturbative approach \cite{Hu2016}, coupled-cluster(CC) \cite{Hagen14}, in-medium similarity renormalization group (IMSRG) \cite{Her16}, and an equation of motion multiphonon method (EMPM) \cite{DeGre17}.
  
The need for restoring the ground-state correlations is also dictated by the fact that HF accounts only for a fraction of the binding energy of all closed-(sub)shell nuclei throughout the periodic table if modern realistic potentials are adopted.
 This emerges blatantly from the results presented in Ref. \cite{DeGre17}. 

In the present work, we adopted the method developed in Ref. \cite{Cat98} and extended to nuclei in Ref. \cite{Papa07} to carry out a systematic study of bulk properties, nuclear responses, and spectra of a large number of closed (sub-)shell  nuclei ranging from $^4$He to $^{208}$Pb. 
 
We use the chiral potential $\Delta$N$^2$LO$_{\mathrm{GO}} $(394) \cite{Jiang20}, incorporating the three-body force, to generate a self-consistent HF basis and then solve the eigenvalue equations  within TDA, RPA, and RRPA. 

To our knowledge, this is the first {\it ab initio} renormalized version of RPA that directly employs the state-of-art two- and three-body interactions without intermediate smoothing steps. As we shall show, it describes within a unified framework  nuclear bulk properties with accuracy comparable to the one achieved by the CC and IMSRG methods and yields, within the limits imposed by  1p-1h space, complete spectra and responses referred to the correlated ground state. Moreover, it covers all sub-shell-closed nuclei across the nuclear chart, including the heavy systems. 

All these results were obtained at a computational cost significantly lower than the one spent by the available IMSRG solver \cite{StroIMSRG}.

Thus, the present approach offers a new efficient manageable tool for a systematic study of low and high energy nuclear properties from first principles.  
It should have, therefore,  an important impact on nuclear structure and  nuclear astrophysics.

{\it The RRPA formalism.} Following the procedure of Refs. \cite{Rowe68a,Cat96,Cat98}  we consider states of the form
 \begin{equation}
 \ket{\nu}  =  Q^\dagger_\nu \ket{0} =
 \sum_{p,h} [X^\nu_{ph} B^\dagger_{ph} - Y^\nu_{ph} B_{ph}] \ket{0}. 
 \end{equation}
Here $B^\dagger_{ph}=  D^{-1/2}_{ph} a^\dagger_{p} a_{h}^{}$ 
denotes a renormalized p-h creation operator with respect to the  HF vacuum, and
\begin{equation}
D_{ph} = n_h - n_p,  
\end{equation}
where $n_r = \bra{0} a^\dagger_r a_r \ket{0}$  are the particle ($r=p$) and hole ($r=h$)
ground-state occupation numbers.
The ground state is obtained by imposing the condition
 \begin{equation}
 Q_\nu \ket{0}  = 0.
 \end{equation}
 In the QBA underlying RPA, $\ket{0}$ is replaced by  the unperturbed HF wavefunction  so that $n_p=0$, $n_h=1$  and $D_{ph}=1$.   

Using the equation of motion method  \cite{Rowe68a}, one obtains
\begin{eqnarray}
\label{Eig}
 \left(
\begin{array}{cc}
\mathcal{A} & \mathcal{B} \\
-\mathcal{B}^{*} & -\mathcal{A}^{*}
\end{array}
\right)
\left(
\begin{array}{c}
X^\nu\\
Y^\nu
\end{array}
\right)  =\hbar \omega_\nu 
\left(
\begin{array}{c}
X^\nu\\
Y^\nu
\end{array}
\right), \label{eq:RPA}
\end{eqnarray}
where 
$\hbar \omega_\nu =  E_\nu - E_0$
and
\begin{align}
    \begin{split}
        \mathcal{A}_{p h, p' h'} &=  \bra{0} [B_{ph},H,B^\dagger_{p'h'}] \ket{0}, \\
        \mathcal{B}_{p h, p' h'}&= -  \bra{0} [B^\dagger_{ph},H,B^\dagger_{p'h'}] \ket{0},
    \end{split}
\end{align}
 having introduced the symmetrized double commutator
 \begin{equation}
 [A,B,C]=\frac{1}{2} ([A,[B,C]] + [[A,B],C]).
 \end{equation} 
 The calculation of the block matrices yields
 \begin{align}
    \begin{split}
    \mathcal{A}_{p h, p' h'}  = &{\cal D}_{php'h'} \left (\epsilon_{pp'} \delta_{hh'}  - \epsilon_{hh'} \delta_{pp'} \right) \\
    &+ D^{1/2}_{ph} D^{1/2}_{p'h'} \bra{hp'} V \ket{ph'} , \\
    \mathcal{B}_{p h, p' h'} =  &D^{1/2}_{ph} D^{1/2}_{p'h'} \bra{hh'} V \ket{pp'},
    \end{split}
 \end{align}
 where ($r=p,h$ and $s=p,h$)
 \begin{align}
 \label{eps}
 {\cal D}_{php'h'} =& \frac{1}{2} ( D^{1/2}_{ph} D^{-1/2}_{p'h'} + D^{-1/2}_{ph} D^{1/2}_{p'h'}),
 \\
 \begin{split}
 \epsilon_{pp'}  =& \bra{p'} t \ket{p} + \sum_r n_r \bra{p'r} V \ket{pr} \\
 &+ \frac{1}{2} \sum_{r,s}  n_{r}^{} n_{s}^{} \bra{p'rs} V \ket{prs}, \end{split} \\
 \begin{split}
\epsilon_{hh'} =& \bra{h} t \ket{h'} + \sum_r n_r \bra{h r} V \ket{h'r} \\
 &+ \frac{1}{2} \sum_{r,s} n_{r}^{} n_{s}^{} \bra{hrs} V \ket{h'rs}.
 \end{split}
\end{align}
The occupation numbers $n_r$ are to be obtained from the one-body density matrix (OBDM)  
\begin{align}
\begin{split}
      \langle 0|a_{p}^{\dag} a_{p'}^{} | 0 \rangle =& \sum_{\nu, \mu, h} \Big[ \delta_{\nu \mu}^{} - \frac{1}{2} \sum_{q,g} D_{qg}^{} X_{qg}^{\mu} X_{qg}^{\nu\,*} \Big] \cdot \\ &\cdot D_{ph}^{1/2} D_{p'h}^{1/2} Y_{ph}^{\nu} Y_{p'h}^{\mu \, *} + \mathcal{O} \big( |Y|_{}^{6}\big),  \\
      \langle 0|a_{h}^{\dag} a_{h'}^{} | 0 \rangle =& \delta_{hh'}^{}  \rangle -       
      \sum_{\nu, \mu, p} \Big[ \delta_{\nu \mu}^{} - \frac{1}{2} \sum_{q,g} D_{qg}^{} X_{qg}^{\mu} \cdot \\ &\cdot X_{qg}^{\nu\,*} \Big] D_{ph}^{1/2} D_{ph'}^{1/2} Y_{ph}^{\nu} Y_{ph'}^{\mu \, *} + \mathcal{O} \big( |Y|_{}^{6}\big). \label{eq:OBDM}
  \end{split}
\end{align}
The above system of nonlinear coupled equations is solved iteratively. Each iteration gives a new OBDM whose diagonalization yields a new set of occupation numbers $n_{r}$ ($r=p,h$) and defines a new natural orbital 
single-particle basis. The iteration ends when the convergence is reached.

The whole process 
implies a double iteration.
One starts with solving the standard RPA eigenvalue problem, obtaining thereby the initial amplitudes $X$ and $Y$. These amplitudes together with the HF occupation numbers are used to determine the density matrices in Eqs.~\eqref{eq:OBDM}. The diagonalization of these latter quantities yields the new occupations numbers and a natural orbital basis to be used for solving again the RPA eigenvalue problem (Eq.~\eqref{eq:RPA}). A new set of RPA amplitudes is obtained thereby. The above procedure is iterated until the OBDM converges to a fixed point.

Final RPA amplitudes define excited states built on top of the correlated ground state and can be used to evaluate the ground-state transition amplitudes of any one-body operator $F$ 
\begin{equation}
\label{F}
\bra{\nu} F \ket{0} = \sum_{p,h} D_{ph}^{1/2} \left( X^{\nu \,*}_{ph} \langle p|F|h \rangle
+ Y^{\nu \,*}_{ph} \langle h|F|p \rangle \right).
\end{equation}
In the limit $D_{ph} \rightarrow 1$, the standard RPA formula is recovered.
 
{\it Numerical implementation and results.}
The Hamiltonian is composed of the intrinsic kinetic energy  $T_{\mathrm{int}}$  and   the chiral potential $ \Delta$N$^2$LO$_{\mathrm{GO}} $(394) \cite{Jiang20}.
 
A HF basis was generated within an harmonic oscillator (HO) space covering all  major shells up to $N_{\mathrm{max}}=14$. Such a basis was adopted to determine the matrix elements of the interaction. The \texttt{NuHamil} numerical code was used for this purpose \cite{Miy23}. 

All matrix elements of the three-body potential up to $N^{(3 )}_{\mathrm{max}} \equiv \mathrm{min(3N_{\mathrm{max}}, 28)}$ are included at the normal ordered two-body (NO2B) level \cite{Miy22}. These are the only ones entering in calculations performed within a p-h configuration space as in our case. Such a p-h basis, with the exception of the HF vacuum, is also free of spurious center of mass admixtures. These are removed by an orthogonalization procedure \cite{DEGREGORIO2021}. 

The ground-state energy is given by
 \begin{equation}
E_0 = E_{\mathrm{HF}} + E_{\mathrm{corr}},
 \end{equation}
where $E_{\mathrm{HF}}$ is the HF term and 
\begin{equation}
 E_{\mathrm{corr}} = -\sum_\nu \hbar \omega_\nu \sum_{p,h}|Y_{ph}|^2
\end{equation}
comes from the correlations.
The neutron ($\tau = \nu$) and  proton ($\tau = \pi$)  square radii  
\begin{equation}
 \label{rp}
\langle r_\tau^2 \rangle
= \frac{1}{N_\tau} \bra{0}\sum_{i=1}^{N_\tau} (\vec{r}_\tau(i) - \vec{R}_{\mathrm{c.m.}})^2 \ket{0} 
\end{equation}
are referred  to the  center of mass (c.m.) in order to minimize the spurious admixtures present in the HF vacuum.
The empirical charge radii are deduced  from $\langle r_\pi^2 \rangle$ through the formula given, for instance,  in Refs. \cite{Hu2016,DeGregorio2017}. 

\begin{figure}[!t]
\includegraphics[width=0.95\columnwidth]{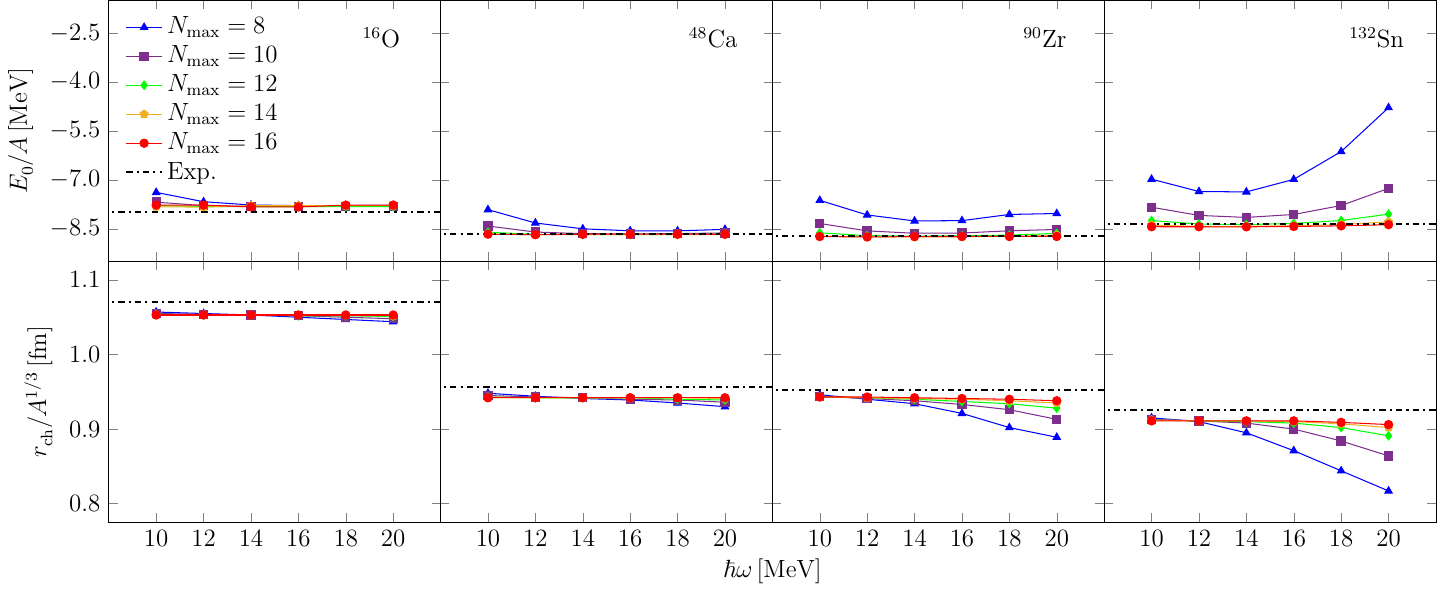}
\caption{(Color online) \label{fig1} Convergence of the RRPA binding energies per nucleon and charge radii versus the HO frequency $\hbar \omega$ for different numbers $N_{\mathrm{max}}$ of HO major shells. The dash-dotted lines indicate the experimental values \cite{NuDat, Angeli13}.} 
\end{figure}

As shown in Fig. \ref{fig1}, the ground-state energies as well as the charge radii become insensitive to the HO frequency only in a space encompassing an increasing number of major shells as we move from light to heavy nuclei. An overall fair agreement between computed and empirical values is attained. For nuclei up to $^{90}$Zr the employed model space is large enough to yield stable results, being their dependence on the HO frequency $\hbar \omega$ very weak. Thus, the  numerical results are presented for a single choice $\hbar \omega = 12$ MeV.  

\onecolumngrid\
\twocolumngrid\

\begin{figure}[ht]
\includegraphics[width=\columnwidth]{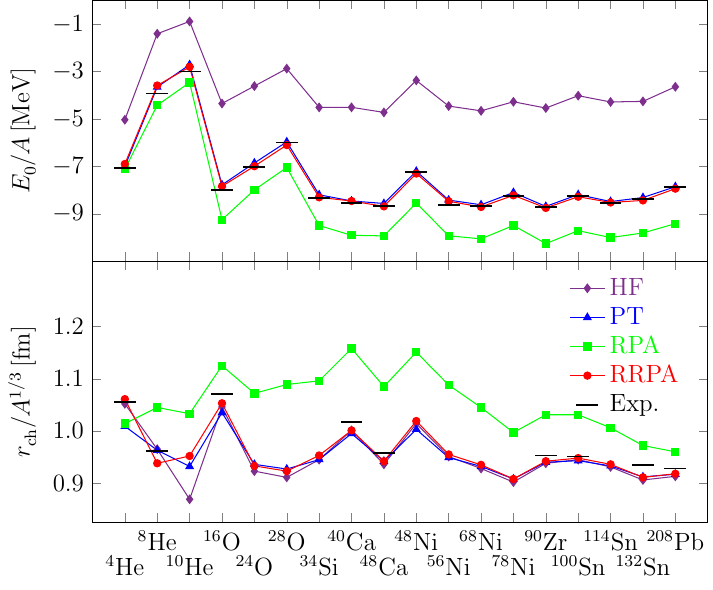}
\caption{(Color online) \label{fig2} Systematic of the HF, PT, RPA, and RRPA binding energies per nucleon and and charge radii versus the empirical values  taken from \cite{NuDat, Angeli13}.} 
\end{figure}

The systematics presented in Fig.~\ref{fig2} prove that HF underestimates severely the binding energy of all studied nuclei. Second-order perturbation theory (PT) refines the calculated energies, significantly reducing discrepancies with the experimental values. On the other hand,  a strong over-binding  is obtained once we add the contribution from the RPA ground-state correlations. The quenching action of the RRPA restores the consistency with the empirical values. 

A more detailed analysis (Table \ref{tab:BE}) shows that RRPA reproduces the binding energies with an overall accuracy comparable to the one achieved within the CC and IMSRG approaches.

\begin{table}[htp]
\begin{center}
\begin{tabularx}{\columnwidth}{YYYYYY}
 \hline 
 \hline
                         &   RRPA     &  PT    &  IMSRG(2)    &     CC      & Exp.   \\ 
 \hline 
  $\,^{16}_{}\mathrm{O}$   &  124.7  &   124.4      &  126.2 & 127.5   &  127.62 \\ 
  $\,^{24}_{}\mathrm{O}$   &  168    &   164        &  167   & 169     &  168.96 \\ 
  $\,^{40}_{}\mathrm{Ca}$  &  338    &   338        &  339   & 346     &  342.05 \\ 
  $\,^{48}_{}\mathrm{Ca}$  &  416    &   410        &  415   & 420     &  416.00 \\ 
  $\,^{78}_{}\mathrm{Ni}$  &  634    &   624        &  629   & 639     &  641.55 \\ 
  $\,^{90}_{}\mathrm{Zr}$  &  773    &   764        &  768   & 782     &  783.90 \\ 
  $\,^{100}_{}\mathrm{Sn}$ &  807    &   793        &  804   & 818     &  825.30 \\ 
  $\,^{132}_{}\mathrm{Sn}$ & 1045    &  1021        &  1026  & 1043    & 1102.84 \\ 
 \hline
 \hline
 \end{tabularx}
 \caption{Binding energies (in MeV) of selected nuclei computed within different theoretical approaches versus experimental data \cite{NNDC}. CC results are taken from \cite{Jiang20}. IMSRG(2) results were obtained with \texttt{imsrg++} code \cite{StroIMSRG}.} 
\label{tab:BE}
\end{center}
\end{table}
\begin{table}[h!]
\begin{center}
\begin{tabularx}{\columnwidth}{YYYY}
 \hline 
 \hline  
    &   RRPA  &  CC   &    Exp.   \\ 
 \hline 
  $\,^{16}_{}\mathrm{O}~~~3^-_1$  &  6.3  &  5.6    &  6.13     \\ 
  $\,^{24}_{}\mathrm{O}~~~2^+_1$  &  3.8  &  3.9    &  4.79     \\ 
  $\,^{48}_{}\mathrm{Ca}~~~2^+_1$  &  4.6  & 4.1    &  3.83     \\ 
  \hline
 \hline
 \end{tabularx}
 \caption{Energies (in MeV) of lowest excited levels calculated within RRPA and CC \cite{Jiang20} methods.}
 \label{tab:levs}
 \end{center}
 \end{table}
 
RRPA restores also the agreement between theoretical and empirical radii overestimated in RPA (Fig.~\ref{fig2}).

\begin{figure}[ht]
\includegraphics[width=0.75\columnwidth]{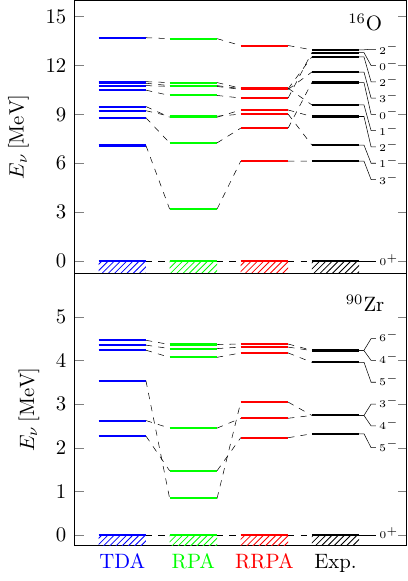}
\caption{(Color online) \label{fig4} Selected low-lying levels of $^{16}$O (top) and $^{90}$Zr (bottom) calculated within TDA, RPA, and RRPA.} 
\end{figure}

Next, we discuss the effect of the renormalization on selected excited states. In RPA, few levels of the $^{16}$O and $^{90}$Zr spectra  are strongly pushed down toward the ground state (Fig. \ref{fig4}). This is a general feature of all investigated nuclei. In some of them, like $^{40}$Ca, the lowest level collapses and the energy becomes imaginary. These instabilities are removed within the RRPA and the consistency  with the experiments is restored. 

 As shown in Table \ref{tab:levs},  the  low lying levels, obtained using the same potential and model space, are reproduced with comparable accuracy in RRPA and  CC \cite{Jiang20}. 

The RRPA monopole energy centroids are also comparable to the corresponding quantities obtained within recent LIT IMSRG/CC approaches \cite{Bona25}.

\begin{figure}[ht]
\includegraphics[width=\columnwidth]{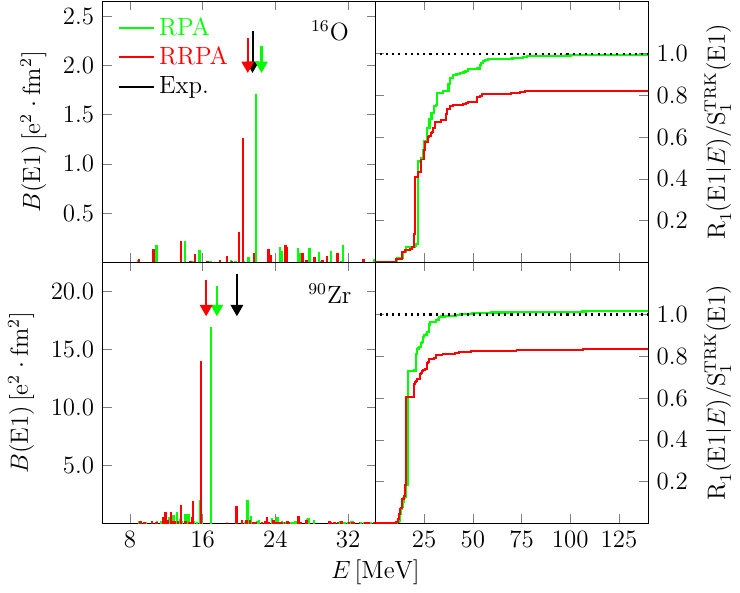}
\caption{(Color online) \label{fig6} RPA versus RRPA reduced $\mathrm{E}1$ strength distributions (left) and the energy-weighted running sums normalized to Thomas-Reiche-Kuhn (TRK) sum rule (right). Arrows indicate the energy centroids calculated as ratio between energy-weighted and non-weighted sums $m_{1}^{}/m_{0}^{}$. Experimental values were adopted from \cite{Goriely-et-al-2019} (Ahrens et al. 1972, Askin et al. 1972 data sets).}
\end{figure}

\begin{figure}[!ht]
\includegraphics[width=\columnwidth]{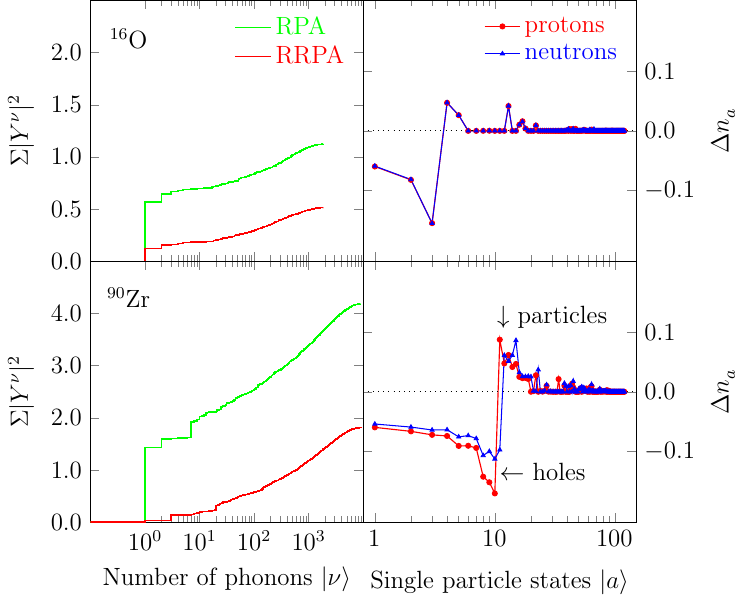}
\caption{(Color online) \label{fig7} Running sum of RPA and RRPA backward amplitudes (left), and deviations of the RRPA from the HF occupation numbers $\Delta n_{a}^{}=n_{a}^{\mathrm{RRPA}}-n_{a}^{\mathrm{HF}}$ (right) in $^{16}$O and $^{90}$Zr. Phonons and single-particle states are ordered according to energy.} 
\end{figure}

Further insight can be obtained from the transition probabilities. The ground-state correlations, accounted for in RRPA. do not further produce any significant fragmentation of the strength with respect to RPA. On the other hand, because of the energy downshift and damping of  the main  E1 peak they cause, the energy-weighted sum rule, preserved in RPA, is sensibly underestimated (Fig. \ref{fig6}).  
This impact is to be attributed to the depletion of few single particle (hole) states since the majority of the states involved in the transitions, being far from the Fermi surface, are empty (full) (Fig.\ref{fig7}).

The depletion of single-particle levels around such a surface strongly affects the low-lying octupole transitions (Table \ref{tab1}). In fact, the $\mathrm{E}3$ ground-state reduced strength of the transition to the 3$^-_1$, practically negligible in TDA, increases dramatically as we move to RPA which systematically overestimates
the measured values. These are sensibly underestimated in RRPA as a consequence of the drastic quenching of the $Y$ amplitudes (Fig.~\ref{fig7}).  

It is not difficult to find the reason of these dramatic changes in going from one approach to the other. Small variations in the $X$ and $Y$ amplitudes are strongly amplified by their mutual interference produced once the transition amplitudes get squared.  

\renewcommand{\arraystretch}{1.2}
\begin{table}[htp]
\begin{center}
\begin{tabularx}{\columnwidth}{YYYYY}
 \hline 
 \hline
     & TDA & RPA & RRPA & Exp. \\ 
 \hline 
  $\,^{16}_{}\mathrm{O}$ & $0.7$ & $2.9$ & $1.2$ & $1.5 \pm 0.1$ \\ 
  $\,^{40}_{}\mathrm{Ca}$ & $5.4$ & $14.8$ & $12.6$ & $20.4 \pm 1.7$ \\ 
  $\,^{48}_{}\mathrm{Ca}$ & $5.7$ & $17.2$ & $9.6$ & $8.3 \pm 0.2$ \\ 
  $\,^{68}_{}\mathrm{Ni}$ & $2.2$ & $21.7$ & $8.0$ & $38.0 \pm 9.0$ \\ 
  $\,^{90}_{}\mathrm{Zr}$ & $27.1$ & $396.6$ & $58.7$ & $108.0 \pm 9.0$ \\ 
  $\,^{114}_{}\mathrm{Sn}$ & $7.0$ & $87.6$ & $34.8$ & $100.0 \pm 12.0$ \\ 
  $\,^{208}_{}\mathrm{Pb}$ & $123.6$ & $773.8$ & $277.7$ & $611.0 \pm 9.0$ \\  
 \hline
 \hline
 \end{tabularx}
\caption{Reduced  strengths $B(\mathrm{E}3, 0_{1}^{+} \rightarrow 3_{1}^{-})$ $\mathrm{[10^3 \cdot  e^2  \cdot fm^6]}$ of the $\mathrm{E}3$ transitions from  the ground to the first octupole states in few selected nuclei. The experimental values are taken from Ref. \cite{Spear-1989}.}
\label{tab1}
\end{center}
\end{table}

{\it Concluding remarks.}
 The limits of RPA emerge definitely from the present systematic based on the use of modern realistic potentials: The overestimation of binding energies and charge radii, the softness of some excited levels toward the ground state.  
 
 The origin of these deficiencies is suggested by the plot in Fig. \ref{fig7}.
 We infer from the too large values of the backward $Y$ amplitudes that the strong ground-state correlations induced  by modern realistic potentials invalidate the QBA underlying RPA. Once this approximation is removed, the amplitudes are damped and the inconsistencies removed.
 
The present approach achieves such a remarkable result  without spoiling the simple  structure of the RPA eigenvalue equations. One needs to  account for the deviations of the single particle occupation numbers from 0 or 1. 
These deviations are quite pronounced in proximity of the Fermi surface and vanish rapidly as we move away from (Fig. \ref{fig7}). Moreover, it is necessary to include the contribution coming from the non-diagonal particle-particle (pp') and hole-hole (hh') (Eqs. \ref{eq:OBDM}) terms entering the OBDMs, as proposed  in Ref. \cite{Cat98}, in order to reproduce the charge radii, strongly overestimated in RPA.

As pointed out already, the RRPA levels and responses are referred to  the true correlated ground state.  A significant link with no-core shell model \cite{Barrett13} and coupled-cluster \cite{Hagen14} is established thereby.
Such a link will be reinforced once the method will be recast in terms of Bogoliubov quasiparticles (qp), a feasible task, so as to cover a large fraction of open shell nuclei.

Even so reformulated, however, RRPA cannot be considered a complete alternative to no-core shell model. Being confined within a p-h or 2qp configuration space, RRPA can account only for a fraction of the energy levels forming the nuclear spectra and is unable to fully satisfy the energy-weighted sum rule.
One needs to enlarge the space so as to include $np-nh$ ($n>1$) or $nqp$  ($n>2$) configurations in order  to cover the full energy spectrum and, hopefully, to reduce the persisting discrepancies between theoretical and experimental responses. 

This upgrade is not at all trivial if we stick on RPA and its extensions. It is more natural to resort to a  closely related and reliable alternative  represented by the EMPM in its qp version \cite{DeGreg16}.

We can therefore conclude that, within the limits of its validity, the present self-consistent RRPA approach offers, from first principles, a simple reliable unified systematic of bulk and spectroscopic properties of finite closed (sub-)shell nuclei at a moderate numerical computational cost, and is especially useful in the regions of medium and heavy nuclei, not easily  accessible to the other well established {\it ab initio}  approaches. Hopefully, it might open a new route, based on the direct use of bar modern potentials, to the large community adopting mean field approaches.

\begin{acknowledgments}
This work  is supported by the Czech Science Foundation (Czech Republic), P203-26-21972S. Computational resources were provided by the e-INFRA CZ project (ID:90254), supported by the Ministry of Education, Youth and Sports of the Czech Republic, and by the ELIXIR-CZ project (ID:90255). 
\end{acknowledgments}
\bibliographystyle{apsrev}
\bibliography{ExtRPA}

@book{rowe10,
  title={Nuclear collective motion: models and theory},
  author={Rowe, David J},
  year={2010},
  publisher={World Scientific}
}

@book{ring04,
  title={The nuclear many-body problem},
  author={Ring, Peter and Schuck, Peter},
  year={2004},
  publisher={Springer Science \& Business Media}
}

@article{Bend03,
  title = {Self-consistent mean-field models for nuclear structure},
  author = {Bender, Michael and Heenen, Paul-Henri and Reinhard, Paul-Gerhard},
  journal = {Rev. Mod. Phys.},
  volume = {75},
  issue = {1},
  pages = {121--180},
  numpages = {0},
  year = {2003},
  month = {Jan},
  publisher = {American Physical Society} 
}

@article{Roca18,
    author = {X. Roca-Maza and N. Paar},
    title = {Nuclear equation of state from ground and collective excited state properties of nuclei },
    journal = {Progress in Particle and Nuclear Physics},
    year = {2018},
    volume = {101},
    pages ={96}
}

@article{Vret05,
    author = {D. Vretenar and A.V. Afanasjev and G.A. Lalazissis and P. Ring},
    title = {Relativistic Hartree–Bogoliubov theory: static and dynamic aspects of exotic nuclear structure},
    journal = {Physics Reports} ,
    year = {2005},
    volume = {409},
    number = {3},
    pages = {101}
}

@article{Paar08,
  title = {Quasiparticle random phase approximation based on the relativistic Hartree-Bogoliubov model},
  author = {Paar, N. and Ring, P. and Nik\ifmmode \check{s}\else \v{s}\fi{}i\ifmmode \acute{c}\else \'{c}\fi{}, T. and Vretenar, D.},
  journal = {Phys. Rev. C},
  volume = {67},
  issue = {3},
  pages = {034312},
  numpages = {15},
  year = {2003},
  month = {Mar},
  publisher = {American Physical Society}
}

@article{Lit23,
    author ={Litvinova, E.},
    title = {On the dynamical kernels of fermionic equations of motion in strongly-correlated media},
    journal = {The European Physical Journal A},
    year = {2023},
    volume= {59},
    number = {12},
    pages = {291}
}

@article{Paar06,
  title = {Collective multipole excitations based on correlated realistic nucleon-nucleon interactions},
  author = {Paar, N. and Papakonstantinou, P. and Hergert, H. and Roth, R.},
  journal = {Phys. Rev. C},
  volume = {74},
  issue = {1},
  pages = {014318},
  numpages = {14},
  year = {2006},
  month = {Jul},
  publisher = {American Physical Society}
}

@article{Papa10,
  title = {Large-scale second random-phase approximation calculations with finite-range interactions},
  author = {Papakonstantinou, P. and Roth, R.},
  journal = {Phys. Rev. C},
  volume = {81},
  issue = {2},
  pages = {024317},
  numpages = {13},
  year = {2010},
  month = {Feb},
  publisher = {American Physical Society} 
}

@article{Herg11,
  title = {Quasiparticle random-phase approximation with interactions from the Similarity Renormalization Group},
  author = {Hergert, H. and Papakonstantinou, P. and Roth, R.},
  journal = {Phys. Rev. C},
  volume = {83},
  issue = {6},
  pages = {064317},
  numpages = {17},
  year = {2011},
  month = {Jun},
  publisher = {American Physical Society}
}

@article{Beau23,
  title = {Zero- and finite-temperature electromagnetic strength distributions in closed- and open-shell nuclei from first principles},
  author = {Beaujeault-Taudi\`ere, Y. and Frosini, M. and Ebran, J.-P. and Duguet, T. and Roth, R. and Som\`a, V.},
  journal = {Phys. Rev. C},
  volume = {107},
  issue = {2},
  pages = {L021302},
  numpages = {6},
  year = {2023},
  month = {Feb},
  publisher = {American Physical Society}
}

@article{Rowe68,
  title = {Methods for Calculating Ground-State Correlations of Vibrational Nuclei},
  author = {Rowe, D. J.},
  journal = {Phys. Rev.},
  volume = {175},
  issue = {4},
  pages = {1283--1292},
  numpages = {0},
  year = {1968},
  month = {Nov},
  publisher = {American Physical Society}
}

@article{Rowe68a,
  title = {Equations-of-Motion Method and the Extended Shell Model},
  author = {Rowe, D. J.},
  journal = {Rev. Mod. Phys.},
  volume = {40},
  issue = {1},
  pages = {153--166},
  numpages = {0},
  year = {1968},
  month = {Jan},
  publisher = {American Physical Society}
}

@article{Jiang20,
title = {Accurate bulk properties of nuclei from $A=2$ to $\ensuremath{\infty}$ from potentials with $\mathrm{\ensuremath{\Delta}}$ isobars},
author = {Jiang, W. G. and Ekstr\"om, A. and Forss\'en, C. and Hagen, G. and Jansen, G. R. and Papenbrock, T.},
journal = {Phys. Rev. C},
volume = {102},
issue = {5},
pages = {054301},
numpages = {8},
year = {2020}
}

@article{ELLIS87,
title = {Particle-hole and particle-particle RPA ground state correlations},
journal = {Nuclear Physics A},
volume = {467},
number = {2},
pages = {173-184},
year = {1987},
author = {P.J. Ellis}}

@article{LENSKE90,
title = {RPA ground state correlations in nuclei},
journal = {Physics Letters B},
volume = {249},
number = {3},
pages = {377-380},
Year = {1990},
author = {H. Lenske and J. Wambach}
 }

@article{Gamba15,
  title = {Subtraction method in the second random-phase approximation: First applications with a Skyrme energy functional},
  author = {Gambacurta, D. and Grasso, M. and Engel, J.},
  journal = {Phys. Rev. C},
  volume = {92},
  issue = {3},
  pages = {034303},
  numpages = {9},
  year = {2015},
  month = {Sep},
  publisher = {American Physical Society}
}

@article{KARA93,
title = {Ground state correlations and charge transition densities},
journal = {Physics Letters B},
volume = {306},
number = {3},
pages = {197-200},
year = {1993},
author = {D. Karadjov and V.V. Voronov and F. Catara}}

@article{CATA94,
	author = {F. Catara and N. {Dinh Dang} and M. Sambataro},
	journal = {Nuclear Physics A},
	number = {1},
	pages = {1-12},
	title = {Ground-state correlations beyond RPA},
	volume = {579},
	year = {1994}}

@article{Cat96,
	author = {Catara, F. and Piccitto, G. and Sambataro, M. and Van Giai, N.},
	journal = {Phys. Rev. B},
	number = {24},
	pages = {17536-17546},
	title = {Towards a self-consistent random-phase approximation for Fermi systems},
	volume = {54},
	year = {1996}}

@article{Wu18,
  title = {Chiral ${\mathrm{NNLO}}_{\mathrm{sat}}$ descriptions of nuclear multipole resonances within the random-phase approximation},
  author = {Wu, Q. and Hu, B. S. and Xu, F. R. and Ma, Y. Z. and Dai, S. J. and Sun, Z. H. and Jansen, G. R.},
  journal = {Phys. Rev. C},
  volume = {97},
  issue = {5},
  pages = {054306},
  numpages = {12},
  year = {2018},
  month = {May},
  publisher = {American Physical Society} 
}

@article{Mina16,
  title = {Estimation of a 2p2h effect on Gamow-Teller transitions within the second Tamm-Dancoff approximation},
  author = {Minato, F.},
  journal = {Phys. Rev. C},
  volume = {93},
  issue = {4},
  pages = {044319},
  numpages = {11},
  year = {2016},
  month = {Apr},
  publisher = {American Physical Society} 
}

@article{Bart21,
  title = {Nuclear ground states in a consistent implementation of the time-dependent density matrix approach},
  author = {Barton, Matthew and Stevenson, Paul and Rios, Arnau},
  journal = {Phys. Rev. C},
  volume = {103},
  issue = {6},
  pages = {064304},
  numpages = {14},
  year = {2021},
  month = {Jun},
  publisher = {American Physical Society} 
}

@article{Hu2016,
	author = {Hu, B. S. and Xu, F. R. and Sun, Z. H. and Vary, J. P. and Li, T.},
	issue = {1},
	journal = {Phys. Rev. C},
	numpages = {11},
	pages = {014303},
	title = {Ab initio nuclear many-body perturbation calculations in the Hartree-Fock basis},
	volume = {94},
	year = {2016}}

@article{DeGre17,
  title = {Ground-state correlations within a nonperturbative approach},
  author = {De Gregorio, G. and Herko, J. and Knapp, F. and Lo Iudice, N. and Vesel\'y, P.},
  journal = {Phys. Rev. C},
  volume = {95},
  issue = {2},
  pages = {024306},
  numpages = {9},
  year = {2017},
  month = {Feb},
  publisher = {American Physical Society} 
}

@article{DEGREGORIO2021,
	author = {G. {De Gregorio} and F. Knapp and N. {Lo Iudice} and P. Vesel{\'y}},
	journal = {Physics Letters B},
	pages = {136636},
	title = {Removal of the center of mass in nuclei and its effects on 4He},
	volume = {821},
	year = {2021}}

@article{DeGregorio2017,
	author = {De Gregorio, G. and Herko, J. and Knapp, F. and Lo Iudice, N. and Vesel\'y, P.},
	issue = {2},
	journal = {Phys. Rev. C},
	month = {Feb},
	numpages = {9},
	pages = {024306},
	publisher = {American Physical Society},
	title = {Ground-state correlations within a nonperturbative approach},
	volume = {95},
	year = {2017}}

@misc{NNDC,
	title = {{\rm Data extracted using the NNDC On-line Data Service from the ENSDF database}}}

@article{Angeli13,
	author = {I. Angeli and K.P. Marinova},
	doi = {https://doi.org/10.1016/j.adt.2011.12.006},
	issn = {0092-640X},
	journal = {Atomic Data and Nuclear Data Tables},
	number = {1},
	pages = {69 - 95},
	title = {Table of experimental nuclear ground state charge radii: An update},
	volume = {99},
	year = {2013}}

@article{Barrett13,
	author = {Barrett, B. R. and Navr\'atil, P. and Vary, J. P.},
	journal = {Progress in Particle and Nuclear Physics},
	pages = {131},
	title = {Ab initio no core shell model},
	volume = {69},
	year = {2013}}

@article{DeGreg16,
	author = {De Gregorio, G. and Knapp, F. and Lo Iudice, N. and Vesel\'y, P},
	journal = {Phys Rev. C},
	pages = {044314},
	title = {Self-consistent quasiparticle formulation of a multiphonon method and its application to the neutron-rich $^{20}\mathrm{O}$ nucleus},
	volume = {93},
	year = {2016}}

@article{Hagen14,
	author = {Hagen, G. and Papenbrock, T. and Hjorth-Jensen, M. and Dean, D. J.},
	journal = {Rep. Prog. Phys.},
	pages = {096302},
	title = {Coupled-cluster computations of atomic nuclei},
	volume = {77},
	year = {2014}}

@article{Schu21,
    title = {Equation of Motion Method for strongly correlated Fermi systems and Extended RPA approaches},
    author = {P. Schuck and D. S. Delion and J. Dukelsky and M. Jemai and E. Litvinova and G. Röpke and M. Tohyama},
    journal = {Physics Reports},
    volume = {929},
    pages = {1-84},
    year = {2021},
    doi = {https://doi.org/10.1016/j.physrep.2021.06.001},
}

@article{Miy23,
	title = {NuHamil},
	author = {Miyagi, T.},
	journal = {Eur. Phys. J. A},
	pages = {150},
	volume = {59},
	year = {2023},
    doi={10.1140/epja/s10050-023-01039-y},
}

@article{Miy22,
  title = {Converged ab initio calculations of heavy nuclei},
  author = {Miyagi, T. and Stroberg, S. R. and Navr\'atil, P. and Hebeler, K. and Holt, J. D.},
  journal = {Phys. Rev. C},
  volume = {105},
  issue = {1},
  pages = {014302},
  numpages = {14},
  year = {2022},
  month = {Jan},
  publisher = {American Physical Society},
  doi = {10.1103/PhysRevC.105.014302},
}

@article{Cat98,
  title = {Self-consistent determination of the one-body density matrix and particle-hole excitations},
  author = {Catara, F. and Grasso, M. and Piccitto, G. and Sambataro, M.},
  journal = {Phys. Rev. B},
  volume = {58},
  issue = {24},
  pages = {16070--16075},
  numpages = {0},
  year = {1998},
  month = {Dec},
  publisher = {American Physical Society},
  doi = {10.1103/PhysRevB.58.16070},
}

@article{Spear-1989,
    title = {Reduced electric-octupole transition probabilities, {B}({E}3; $0_{1}^{+} \rightarrow 3_{1}^{-}$), for even-even nuclides throughout the periodic table},
    journal = {Atomic Data and Nuclear Data Tables},
    volume = {42},
    number = {1},
    pages = {55-104},
    year = {1989},
    issn = {0092-640X},
    author = {R. H. Spear},
    doi = {https://doi.org/10.1016/0092-640X(89)90032-6},
}

@article{Papa07,
  title = {Nuclear collective excitations using correlated realistic interactions: The role of explicit random-phase approximation correlations},
  author = {Papakonstantinou, P. and Roth, R. and Paar, N.},
  journal = {Phys. Rev. C},
  volume = {75},
  issue = {1},
  pages = {014310},
  numpages = {11},
  year = {2007},
  month = {Jan},
  publisher = {American Physical Society},
  doi = {10.1103/PhysRevC.75.014310},
}

@misc{NuDat,
  title = {National Nuclear Data Center database},
  url={https://www.nndc.bnl.gov/nudat3/}
}

@article{Goriely-et-al-2019,
   title={Reference database for photon strength functions},
   volume={55},
   ISSN={1434-601X},
   number={10},
   journal={The European Physical Journal A},
   publisher={Springer Science and Business Media LLC},
   author={Goriely, S. and Dimitriou, P. and Wiedeking, M. and Belgya, T. and Firestone, R. and Kopecky, J. and Krtička, M. and Plujko, V. and Schwengner, R. and Siem, S. and Utsunomiya, H. and Hilaire, S. and Péru, S. and Cho, Y. S. and Filipescu, D. M. and Iwamoto, N. and Kawano, T. and Varlamov, V. and Xu, R.},
   year={2019},
   doi={10.1140/epja/i2019-12840-1}
}

@article{Her16,
title = {The In-Medium Similarity Renormalization Group: A novel \textit{ab initio} method for nuclei},
journal = {Physics Reports},
volume = {621},
pages = {165-222},
year = {2016},
issn = {0370-1573},
author = {H. Hergert and S.K. Bogner and T.D. Moris and A. Schwenk and K. Tsukiyama},
}

@article{Rai19,
  title = {Nuclear electromagnetic dipole response with the self-consistent Green's function formalism},
  author = {Raimondi, Francesco and Barbieri, Carlo},
  journal = {Phys. Rev. C},
  volume = {99},
  issue = {5},
  pages = {054327},
  numpages = {13},
  year = {2019},
  month = {May},
  publisher = {American Physical Society},
  doi = {10.1103/PhysRevC.99.054327}
}

@misc{StroIMSRG,
  author = {S.R.Stroberg},
  publisher = {GitHub},
  journal = {GitHub repository},
  howpublished = {\url{https://github.com/ragnarstroberg/imsrg}},
}

@article{Bona25,
    author = {F. Bonaiti and A. Porro and S. Bacca and A. Schwenk and  A. Tichai},
    title = {Ab initio calculations of monopole sum rules: From finite nuclei to infinite nuclear matter},
    journal ={arXiv:2511.15836}, 
    year = {2025},
}
\end{document}